\documentclass[12pt]{article}
\usepackage{graphicx}

\newcommand\pubnumber{XXXXX}
\newcommand\pubdate{\today}

\def\raleigh{Department of Physics,\\NC State University,\\Raleigh, NC 27695, USA}
\def\support{\footnote{Work supported by the Department of Energy Office of Nuclear Physics, Award number DE-SC0004786}}

\def\Title#1{\begin{center} {\Large #1 } \end{center}}
\def\Author#1{\begin{center}{ \sc #1} \end{center}}
\def\Address#1{\begin{center}{ \it #1} \end{center}}

\newcommand\pubblock{\rightline{\begin{tabular}{l} \pubnumber\\
         \pubdate  \end{tabular}}}
\newenvironment{Abstract}{\begin{quotation}  }{\end{quotation}}
\newenvironment{Presented}{\begin{quotation} \begin{center} 
             PRESENTED AT\end{center}\bigskip 
      \begin{center}\begin{large}}{\end{large}\end{center} \end{quotation}}

\def\beq{\begin{equation}}
\def\eeq#1{\label{#1}\end{equation}}
\def\eeqn{\end{equation}}

\def\beqa{\begin{eqnarray}}
\def\eeqa#1{\label{#1}\end{eqnarray}}
\def\eeqan{\end{eqnarray}}

\let\bar=\overbar

\def\Dslash{\not{\hbox{\kern-4pt $D$}}}
\def\dslash{\not{\hbox{\kern-2pt $\del$}}}

\def\msb{{\bar{\ssstyle M \kern -1pt S}}}

\begin{document}
\begin{titlepage}
\pubblock

\vfill
\Title{The Physics Of Supernova Neutrino Oscillations}
\vfill
\Author{James P.\  Kneller \support}
\Address{\raleigh}
\vfill
\begin{Abstract}
On February 23, 1987 we collected 24 neutrinos from the explosion of a blue super-giant star in the Large Magellanic Cloud confirming the basic paradigm of core-collapse supernova. During the many years we have been waiting for a repeat of that momentous day, the number and size of neutrino detectors around the world has grown considerably. If the neutrinos from the next supernova in our Galaxy arrive tomorrow we shall collect upwards of tens of thousands of events and next generation detectors will increase the amount of data we collect by more than an order of magnitude. But it is also now apparent that the message is much more complex than previously thought because many time, energy and neutrino flavor dependent features are imprinted upon the signal either at emission or by the passage through the outer layers of the star. These features arise due to the explosion dynamics, the physics of nuclei at high temperatures and densities, and the properties of neutrinos. In this proceedings I will present some aspects of the physics of supernova neutrino oscillations and what we should expect to observe when the neutrinos from the next Galactic supernova (eventually) arrive.
\end{Abstract}
\vfill
\begin{Presented}
CIPANP2015, Vail, CO. May 19-24, 2015
\end{Presented}
\vfill
\end{titlepage}
\def\thefootnote{\fnsymbol{footnote}}
\setcounter{footnote}{0}

\section{Introduction}

Although supernovae of all three principle types - thermonuclear, core-collapse and pair-instability - emit neutrinos, 
the number emitted by core-collapse supernovae far outshines the other two types with around $10^{58}$ neutrinos emitted in a period of $\sim$10 seconds.
A core collapse supernova at a distance of $10\;{\rm kpc}$ from Earth will produce of order $10^{4}$ neutrino events in the SuperKamiokande detector. 
Such a signal contains a wealth of information about the details of the explosion that would allow us to test our current understanding of the supernova
phenomenon, as well as information about the neutrino itself which remains an elusive particle whose properties are notoriously difficult to determine. 
For a recent review of supernova neutrinos the reader is referred to the excellent article by Scholberg \cite{2012ARNPS..62...81S}\index{2012ARNPS..62...81S}.

The neutrino emission from a core collapse supernova can be roughly divided into four epochs: the precollapse emission, the collapse epoch, the accretion epoch, and the cooling epoch.

{\bf The precollapse}:
All stars emit neutrinos of course but it is only when a massive star starts to burn Carbon that the neutrino emission becomes larger than the photon emission. Indeed it is the fact that so many neutrinos are produced and escape from the core of such stars that leads to the rapidly decreasing lifetimes of the late stages of stellar evolution. By the time stars reach Silicon burning the neutrino luminosity can be orders of magnitude larger than the photon emission. The neutrinos emitted during Silicon burning from a nearby nearby star ( d $<$ 1kpc ) prior to collapse are detectable in SuperKamiokande and KamLAND a few days to weeks before the explosion
\cite{Kato,Asakura} \index{Kato,Asakura}.

{\bf The collapse}:
The collapse of the core is triggered by electron capture and/or photo disassociation of the nuclei. Once initiated the core contracts from a radius of roughly $r\sim 1000\;{\rm km}$ down to $r\sim 20\;{\rm km}$ 
in a period of order $\sim 200$ milliseconds. The collapse is halted at this point due to nucleon repulsion and the gravitational binding energy released is of order $10^{53}\;{\rm erg}$. The collapse compresses the core beyond its equilibrium and so it rebounds sending a pressure wave outwards into the star. At the point where the velocity of the matter in this pressure wave becomes greater than the local sound speed, the pressure wave steepens into a shock wave. Electron neutrino emission dominates over the emission of all other flavors during this epoch and this is also the phase where the neutronization burst occurs. Surprisingly it is found that the neutrino emission is largely independent of many details one might expect to play a role. For example, the neutronization burst possesses little sensitivity to the equation of state of dense nuclear matter and the progenitor mass \cite{Mayle,OConnor2013} \index{Mayle,OConnor2013}. 

{\bf The accretion (pre-explosion) phase}: The supernova now enters the accretion phase which can be divided into two parts:
the early accretion phase up to $\sim 100$ s post bounce during which the shock wave propagates out to a distance of $\sim 200\;{\rm km}$ or so before stalling, and the later phase, which may last for several hundred milliseconds, 
until the shock is revived. During the second phase mass accretion is channeled into downflows and one observes significant asphericity in the form of convection and/or a standing accretion shock instability (SASI) \cite{2003ApJ...584..971B}\index{2003ApJ...584..971B}. The luminosities of electron neutrino and antineutrinos are now approximately equal. In both phases of the accretion epoch it is found the more compact the star the higher the accretion rate which leads to higher neutrino luminosities and average energies \cite{OConnor2013,2013ApJ...767L...6B} \index{OConnor2013,2013ApJ...767L...6B}: what changes during the second phase of the accretion epoch is that the neutrino emission becomes aspherical. 
In particular, it is observed in simulations that the SASI introduces luminosity fluctuations of order $10\%$ and periodicity of $\sim 10$ milliseconds \cite{2014ApJ...788...82M}\index{2014ApJ...788...82M}. 
As rapid as these may appear, the time resolution of the IceCube detector is of order $1\;{\rm ms}$ which means IceCube can be 
used to observe the SASI in the neutrino signal \cite{Lund2010,Lund2012} \index{Lund2010,Lund2012}.
These fluctuations have the obvious consequence that there is now a line-of-sight dependence during this epoch of the neutrino signal but interestingly an additional observation was 
recently made by Tamborra \emph{et al.} \cite{Tamborra} \index{Tamborra} that the asphericity can be quasi stable. Dubbed the Lepton-number Emission Self-sustained Asymmetry (LESA), Tamborra \emph{et al.} found there was 
a significant dipole - of order $10\%$ - in the $\nu_e$ and $\bar{\nu}_e$ fluxes from their 3D simulation. Only the $\nu_e$ and $\bar{\nu}_e$ fluxes exhibited this pattern, the heavy lepton 
neutrino emission was almost spherically symmetric. The dipole appeared to emerge stochastically but once formed, the orientation of the dipole was stable 
for 100s of milliseconds even through periods where a SASI was present. If confirmed in further studies, there may be many consequences of the LESA that will need to be explored especially 
the neutrino flavor conversion due to self interaction effects. 

{\bf The cooling phase}:
Finally, once the shock is revived and the mantle of the star moving outwards, the neutrino emission is simply due to the cooling of the hot proto-neutron star over 
a period of $10-20$ seconds or so \cite{Fischer}\index{Fischer}. The neutrino fluxes are expected to return to spherical symmetry but it is presently unclear if 
indeed that eventually occurs because multi-dimensional simulations which follow explosions to times greater than $t\sim 1\;{\rm s}$ post 
bounce often find significant accretion onto the proto-neutron at late times \cite{2014ApJ...788...82M}\index{2014ApJ...788...82M}.

\section{Flavor transformation}
The neutrino spectra emitted at the neutrinosphere are distorted as they propagate to detectors here on Earth. A number 
of effects need to be accounted for if one wishes to simulate a burst signal: neutrino self interactions, a dynamic MSW effect, turbulence, decoherence, and Earth matter. 
Not all effects are present at all epochs of the neutrino emission: neutrino self interactions, the dynamic MSW effect and turbulence are only expected 
to appear in the cooling phase of the emission and not during the other periods. 
And of the five effects listed, the Earth matter effect, decoherence and dynamic MSW effect are all well understood from theory and from numerical studies. 
The same cannot be said of neutrino self interactions which have proven difficult to understand from a theoretical standpoint and demanding to compute numerically. 
For the case of turbulence, there has been recent progress on the theory side - which we highlight later - and the present barrier is the 
lack of suitable numerical simulations to calibrate expectations. 

In order to compute the neutrino flavor transformation one must solve for the evolution matrix $S$ 
which relates the neutrino state at a given position $r$ to the initial state at the neutrinosphere. 
$S$ obeys a differential equation
\begin{equation}
\imath \frac{dS}{d\lambda} = H\,S \label{eq:1}
\end{equation}
where $H$ is the Hamiltonian and $\lambda$ is an affine parameter. Once we have found $S$, the probability that an initial state $\nu_j$ is 
detected as state $\nu_i$ at r is simply the square amplitude of the ij element of $S$ i.e.\  $P_{ij} = |S_{ij}|^2$. With only Standard Model physics, the Hamiltonian $H$
is composed of three terms: the vacuum (kinetic energy) term, a contribution $H_{m}$ if the neutrino is propagating through matter \cite{M&S1986,Wolfenstein1977}\index{M&S1986,Wolfenstein1977}, and a third term 
due to neutrino self-interactions \cite{Duan,Raffelt,2007JCAP...12..010F} \index{Duan,Raffelt,2007JCAP...12..010F}. 

When we put together neutrino flavor transformation with the physical inputs to the emission we find the neutrino signal from a Galactic supernova 
is sensitive to both unknown neutrino properties, the properties of nuclei in hot/dense matter, and the details of the core-collapse supernova mechanism. 
For example, it has been shown that one may infer such neutrino properties as: the neutrino mass hierarchy, the number of $\nu$ flavors, neutrino self-interaction effects, 
MSW and turbulence effects, neutrino non-standard interactions, neutrino magnetic moments, and even the SUSY contribution to the neutrino refractive index. 
Similarly the signal is affected by: the progenitor and structure, the neutrino opacities, the nuclear Equation of State, the shock position / velocity, 
the SASI, the stalled shock duration, and many other details. There is not space here to discuss all these different dependencies so 
we choose to focus upon the case of turbulence and the possibility that the neutrinos may be sensitive to Beyond Standard Model physics in the form of non-standard neutrino interactions with matter.

\begin{figure}[htb]
\centering
\includegraphics[height=3in]{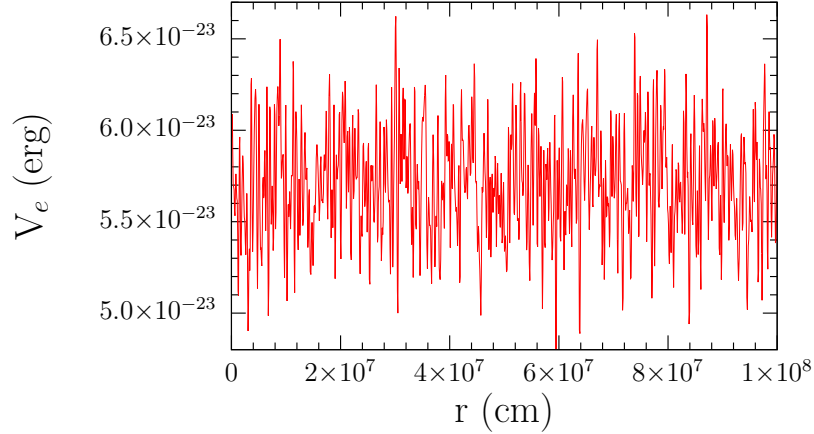}
\caption{A realization of turbulence superposed upon a constant density. The figure is taken from Patton, Kneller \& McLaughlin \cite{Patton2014}\index{Patton2014} and the 
reader is referred to that paper for further details.}
\label{fig:1}
\end{figure}
\begin{figure}[htb]
\centering
\includegraphics[height=3in]{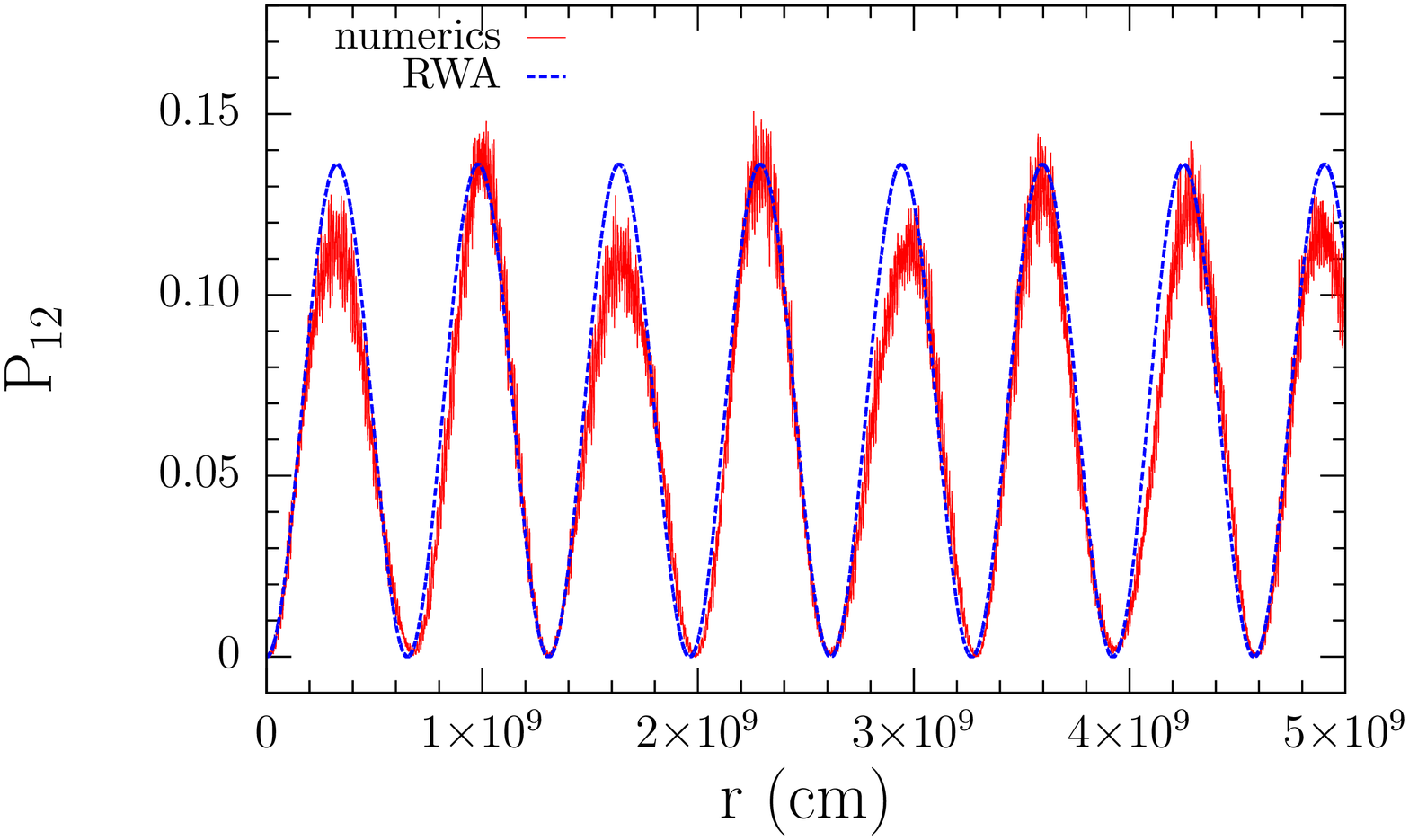}
\caption{The numerical and Rotating Wave Approximation solution for the transition probability $P_{12}$ as a function of distance through the turbulent shown in figure (\ref{fig:1}). 
The figure is taken from Patton, Kneller \& McLaughlin \cite{Patton2014}\index{Patton2014}. }
\label{fig:2}
\end{figure}
\section{Stimulated transitions}

In the presence of matter the neutrinos gain a potential energy due to coherent forward scattering upon the electrons in medium. 
This potential is of the form $V = \pm \sqrt{2}\,G_F\, \rho\,Y_e\, /\, m_N$ where $\rho$ is the mass density, $Y_e$ is the electron fraction and $m_N$ is the nucleon mass. 
The combination $\rho\, Y_e / m_N$ is equivalent to the electron density. 
This potential is the `ee' element of $H_{m}$ and the plus sign is used for the neutrinos, the minus sign for antineutrinos. 
If turbulent density fluctuations of the medium are present then they affect the neutrinos via $\rho$. It would appear that the most obvious way to study 
the effect of turbulence upon the neutrinos is to simply extract from a simulation a profile for $\rho$ and plug it into equation (\ref{eq:1}) and integrate. 
However that option is not possible at the present time because we do not posses suitable high resolution, long duration multi-d simulations 
which can be used. We are thus forced to adopt density profiles from turbulence free 1D hydrodynamical supernova simulations and add turbulence to them. This can be done 
by multiplying $\rho$ from the 1D simulation by $1+F(r)$ where $F(r)$ is a random field. Typically the random field is chosen to be Gaussian with an rms amplitude $C_{\star}$ and a power spectrum that is an 
inverse power law. The random field can be realized with a Fourier series of the form
\begin{equation}\label{eq:F1D}
F(r)\propto C_{\star}\,\sum_{a=1}^{N_q}\,\sqrt{V_{a}}\left\{ A_{a} \cos\left(q_{a}\,r\right) + B_{a} \sin\left(q_{a}\,r\right) \right\}. 
\end{equation}
Here the sets $\{A\}$, $\{B\}$ are the amplitudes of the modes and the wavenumbers are $\{q\}$. An example of a realization of turbulence, taken from \cite{Patton2014} \index{Patton2014}, for the case of a constant 
background profile is shown in figure (\ref{fig:1}).

One may think that such a chaotic, random potential precludes any possibility of predicting the effect of the turbulence even if all the amplitudes and wavenumbers of the Fourier modes were known. 
That is \emph{not} the case, as shown by \cite{Patton2014,Patton2015} \index{Patton2014,Patton2015}. 
If one treats the turbulent component of $H_{m}$ as a perturbation and uses the machinery of time dependent perturbation theory, one finds it is possible to derive an analytic solution for the neutrino evolution after making the `Rotating Wave Approximation' (RWA). The turbulence acts upon the neutrino similar to how radiation acts upon an atom: the neutrino `picks out' the Fourier modes in the turbulence which have frequencies which match the splitting between pairs of eigenvalues of the unperturbed Hamiltonian. For the case of two neutrino flavors the solution for the transition probability between the matter states is particularly simple:
$P_{12} = \frac{|\kappa|^2}{Q^2} \sin^2(Qr)$. The quantities $\kappa$ and $Q$ are analytic functions of the amplitudes $\{A\}$, $\{B\}$ and wavenumbers $\{q\}$. In figure (\ref{fig:2}) we show how well this 
analytic solution matches the numerical solution for the realization of turbulence shown in figure (\ref{fig:1}). 
The good agreement between the theory and numerical results allows us to better understand turbulence in supernovae and even make predictions when we vary 
the parameters describing the turbulence \cite{Patton2015} \index{Patton2015}. The issue now becomes what exactly those parameters are both in nature and in high definition, long duration, multi-dimensional simulations.

\section{BSM physics with supernova neutrinos}
There is also the potential that supernova neutrinos can be used for studying Beyond Standard Model (BSM) physics if we start modifying the Hamiltonian governing neutrino evolution. 
For example \cite{2007PhRvD..76e3001E,2010PhRvD..81f3003E}\index{2007PhRvD..76e3001E,2010PhRvD..81f3003E}: consider a case where we include neutrino self interactions $H_{SI}$ into $H$ and a more general coupling of neutrinos to matter of the form
\begin{equation}
H_{m} = \pm \frac{\sqrt{2}\,G_F \rho}{m_N} \left[ (1+\epsilon_e)\,Y_e + \epsilon_u\,(1+Y_e) + \epsilon_d (2-Y_e) \right]
\end{equation}
The matrices $\epsilon_e$, $\epsilon_u$ and $\epsilon_d$ describe non-standard interactions (NSI) of neutrinos with electrons, up quarks, and down quarks respectively. If we play with the entries in these matrices then we find it is possible to generate a matter-neutrino resonance (MNR). A MNR occurs when the diagonal elements of the total Hamiltonian become equal \emph{and} the neutrinos convert in such a way the resonance condition is maintained due to feedback via the self-interaction Hamiltonian \cite{Malkus2012,Malkus2014} \index{Malkus2012,Malkus2014}.
This kind of resonance was first seen in studies looking at neutrino emission when two neutron stars merge and occurs because there can be an excess of antineutrino emission over neutrino in this scenario. The antineutrino excess drives the self-interaction potential negative. In supernovae an excess of antineutrino emission is not possible so, with only standard model physics, a MNR cannot occur. But with NSI it becomes possible because the matter Hamiltonian can become negative if the $\epsilon$'s are chosen appropriately. In fact in preliminary calculations it is observed the MNR often occurs before the usual self-interaction instabilities. The consequences for the neutrino signal and the nucleosynthesis of NSI for supernova neutrinos are currently being investigated \cite{Vaananen}\index{Vaananen}. 

\section{Summary}
Studying the neutrino in terrestrial experiments is a slow business. The neutrino is so ephemeral that experimenters must push the boundaries of technology and their patience in order to measure its physical properties. But it is nature which pushes the envelope in the case of core-collapse supernovae where one finds densities, temperatures, magnetic fields etc.\ which far exceed anything we can produce here on Earth. In such an environment neutrinos are no longer ghosts, they become important components of the system. A laundry list of properties of the neutrino alter the dynamics of the supernova and changes the signal we expect from the next supernova in our Galaxy. This list includes both familiar standard model physics - such as the mass hierarchy - and BSM physics - such as non-standard interactions. 
Decoding the signal will be challenging but such sensitivity to the neutrino means it would be wise for us to take advantage of this immense potential for probing its properties (and to observe the inner workings of the explosion) by preparing for the day when the neutrinos from the next Galactic supernova (eventually) arrive.


\begin{thebibliography}{99}


\bibitem{2012ARNPS..62...81S} K.\ Scholberg, Annual Review of Nuclear and Particle Science {\bf 62} 81 (2012)

\bibitem{Kato} C.\ Kato \emph{et al.}, arXiv:1506.02358
\bibitem{Asakura} K.\ Asakura \emph{et al.}, arXiv:1506.01175

\bibitem{Mayle} Mayle, J.\ Wilson,\& D.\ Schramm, ApJ {\bf 318} 288 (1987)

\bibitem{OConnor2013} E.\ O'Connor \& C.\ Ott, ApJ {\bf 62} 126 (2013)

\bibitem{2013ApJ...767L...6B}  S.\ W.\ Bruenn \emph{et al.} ApJL {\bf 767} L6 (2013)
\bibitem{2003ApJ...584..971B} J.\ M.\ Blondin, A.\ Mezzacappa \& C.\ DeMarino, ApJ {\bf 584} 971 (2003)

\bibitem{2014ApJ...788...82M} B.\ M{\"u}ller, \& H.-T.\  Janka, ApJ {\bf 788} 82 (2014)

\bibitem{Lund2010} T.\ Lund \emph{et al.}, PRD {\bf 82} 063007 (2010)
\bibitem{Lund2012} T.\ Lund \emph{et al.}, PRD {\bf 86} 105031 (2012)

\bibitem{Tamborra} I.\ Tamborra \emph{et al.},  ApJ {\bf 792} 96 (2014)

\bibitem{Fischer} T.\ Fischer \emph{et al.}, A\&A {\bf 517} 80 (2010)

\bibitem{M&S1986} S.\ P.\ Mikheev and A.\ I.\ Smirnov, Nuovo Cimento C {\bf 9} 17 (1986) 
\bibitem{Wolfenstein1977} L.\ Wolfenstein, PRD {\bf 17} 2369 (1978)

\bibitem{Duan} H.\ Duan et al., PRL {\bf 97} 241101  (2006)
\bibitem{Raffelt} G.\ Raffelt \& A.\ Smirnov, PRD {\bf 76} 081301 (2007)
\bibitem{2007JCAP...12..010F} G.\ Fogli, G., \emph{et al.}, JCAP {\bf 12} 010 (2007)

\bibitem{Patton2014} K.\ Patton, J.\ P.\ Kneller \& G.\ C.\ McLaughlin, PRD {\bf 89} 073022 (2014)
\bibitem{Patton2015} K.\ Patton, J.\ P.\ Kneller \& G.\ C.\ McLaughlin, PRD {\bf 91} 025001 (2015)


\bibitem{2007PhRvD..76e3001E} A.\ Esteban-Pretel, R.\ Tom{\`a}s \& J.\ W.\ F.\ Valle, PRD {\bf 76} 053001 (2007)
\bibitem{2010PhRvD..81f3003E} A.\ Esteban-Pretel, R.\ Tom{\`a}s \& J.\ W.\ F.\ Valle, PRD {\bf 81} 063003 (2010)
\bibitem{Vaananen} D.\ V\"{a}\"{a}n\"{a}nen \emph{et al.}, in preparation

\bibitem{Malkus2014} A.\ C.\ Malkus, A.\ Freidland \& G.\ C.\ McLaughlin, arXiv:1403.5797
\bibitem{Malkus2012} A.\ C.\ Malkus et al, PRD {\bf 86} 085015 (2012)




\end{thebibliography}
\end{document}